\documentclass[twocolumn,prb,aps,floatfix,epsf,psfig,
superscriptaddress,amssymb,longbibliography]{revtex4-2}
\usepackage{graphicx}
\usepackage{amsmath}
\usepackage{physics}
\usepackage[colorlinks=true, citecolor=blue, urlcolor=blue, linkcolor=blue,bookmarks=false,hypertexnames=true]{hyperref} 

\begin{document}
\graphicspath{}

\title{Interacting holes in gated $\text{WSe}_\text{2}$ quantum dots}

\author{Daniel Miravet}
\affiliation{Department of Physics, University of Ottawa, Ottawa, Ontario, Canada K1N 6N5}

\author{Abdulmenaf Alt{\i}nta\c{s}}
\affiliation{Department of Physics, University of Ottawa, Ottawa, Ontario, Canada K1N 6N5}

\author{Alina Wania Rodrigues}
\affiliation{Department of Physics, University of Ottawa, Ottawa, Ontario, Canada K1N 6N5}

\author{Maciej Bieniek}
\affiliation{Institut f\"{u}r Theoretische Physik und Astrophysik,
Universit\"{a}t W\"{u}rzburg, 97074 W\"{u}rzburg, Germany}
\affiliation{Institute of Theoretical Physics, Wroc\l aw University of Science and Technology, Wybrze\.ze Wyspia\'nskiego 27, 50-370 Wroc\l aw, Poland}

\author{Marek Korkusinski}
\affiliation{Department of Physics, University of Ottawa, Ottawa, Ontario, Canada K1N 6N5}
\affiliation{Security and Disruptive Technologies,
National Research Council, Ottawa, K1A0R6, Canada }

\author{Pawe\l \ Hawrylak}
\affiliation{Department of Physics, University of Ottawa, Ottawa, Ontario, Canada K1N 6N5}

\date{\today}

\begin{abstract}
We develop here a theory of  the electronic properties of a finite number of valence holes in gated $\text{WSe}_\text{2}$ quantum dots, considering the influence of spin, valley, electronic orbitals, and many-body interactions. The single-particle wave functions are constructed by combining the spin-up and down states of the highest valence bulk bands employing a multi-million atom ab-initio based tight-binding model solved in the wave-vector space, allowing to study up to 100 nm radius quantum dots atomistically. The effects of the many-body interactions are determined using the configuration interaction (CI) technique, applied up to   $N = 6$ holes occupying up to 6 electronic shells with 42 orbitals. Our results show that N=2 holes are in valley and spin  anti-ferromagnetic ground state, independent of the  interaction strength and the quantum dot size. However, we predict that higher number of holes  can undergo a transition to spontaneously broken symmetry valley and spin polarized ferromagnetic phases, highlighting the interplay between the many-body effects and the quantum dot lateral size and confining potential depth.
\end{abstract}

\maketitle

\setcounter{secnumdepth}{3} 
\section{Introduction}
Single layers of transition metal dichalcogenides (TMDCs) have emerged as an interesting class of materials with rich spectrum of electronic,  optical and magnetic properties owing to the interplay between strong electron-electron interactions and their low-dimensional nature. 
The fabrication of TMDCs at the few atomic layer scale has unlocked new directions in the low-dimensional physics \cite{splendiani2010emerging,scrace2015magnetoluminescence,roch2019spin,van2019virtual,komsa2012two,boddison2021gate,Boddison20231DChannel,mueller2018exciton,schneider2018two,kadantsev2012electronic,conley2013bandgap,Manzeli_Kis_2017}. In contrast with monolayer graphene, TMDCs exhibit a band gap, which makes them attractive for optoelectronic applications \cite{CastroNeto_Geim_2009,Mak_Heinz_2010,Manzeli_Kis_2017}. 

TMDCs unit cell consists of a metal atom (Mo or W) and a chalcogenide dimer ($\text{S}_\text{2}$, $\text{Se}_\text{2}$ or Te$_2$), with the electronic and optical properties primarily governed by the $d$ orbitals of the metal atoms \cite{altintacs2021spin,bieniek2020effect,bieniek2018band}. The presence of strongly localized $d$ orbitals of the transition metal atoms, coupled with the reduced screening in two dimensions, leads to pronounced correlation effects, which in turn give rise to insulating behavior and the formation of excitons with large binding energies \cite{OtsukaCriticalityMott2016,BieniekExcitons2022,WuExcitonMoS2_2015,BorgesexcitonQ2023}.

Additionally, the spin-orbit coupling present in TMDCs plays an important role, leading to the realization of unique spin-dependent phenomena. Due to the conservation of time-reversal symmetry, valleys K and -K exhibit opposite spin splittings, resulting in a spin-valley locking effect. In particular, $\text{WSe}_\text{2}$ stands out due to its large spin-orbit splitting of approximately 500 meV in the valence band, distinguishing it from other TMDC compounds \cite{le2015spin,alidoust2014observation}.  Moreover, it is worth noting that while the bottom of the conduction band in valleys K and -K is predominantly composed of $m_d = 0$ orbitals, the top of the valence band in valleys K and -K is primarily composed of $m_d = -2$ and $m_d = 2$ orbitals, respectively \cite{bieniek2020effect}. This orbital asymmetry between the valleys can give rise to novel features in hole systems, leading to intriguing phenomena and potential applications in spintronics and quantum information processing \cite{altintacs2021spin}.

The progress in graphene and TMDC materials led to the creation of lateral quantum dots (QDs) by the application  of an electrostatic confinement  \cite{Guclu_Hawrylak_2014,Guttinger_Ensslin_2010,McGuire_2016,Wang_Ruffieux_2017,Wang_Kim_2018,Pisoni_Ensslin_2018,Volk_Stampfer_2011,Allen_Yacoby_2012,Eich_Ensslin_2018,Kurzmann_Ihn_2019,freitag2016electrostatically,YasserGrapheneQD2023}. Application of lateral metallic gates  results in the lateral electrostatic confinement of electrons or holes, resulting in the creation of atomic-like states \cite{altintacs2021spin,bieniek2020effect, Eich_Ensslin_2018,freitag2016electrostatically,Kurzmann_Ihn_2019,BrotonsGisbert_Gerardot_2019,Lu_Srivastava_2019,Chakraborty_Vamivakas_2018,Zhang_Guo_2017,bhandari2018imaging,chen2018magnetic,LudkaQD2020,boddison2021gate,JarekPRA2021}.

In this study we develop  a theory of  the electronic properties of finite number of valence holes in gated $\text{WSe}_\text{2}$ quantum dots, considering the influence of spin, valley, electronic orbitals, and many-body interactions  \cite{song2015gate,LudkaQD2020,boddison2021gate}. The shell spacing $\omega$ of the atomic-like quantum dot levels strongly depends on the lateral size and depth of confining potential embedded in $\text{WSe}_2$ computational box size. Here, we investigate computational boxes containing millions of atomic orbitals, allowing for a comprehensive exploration of the effects of strong interactions on the ground and excited states of valence holes confined by lateral potentials to a quantum dot of definite lateral size and potential depth.

\hfill 

The paper is structured as follows. In Section \ref{sect:model}, we provide a comprehensive description of the single-particle states in the gated $\text{WSe}_\text{2}$ quantum dot, along with an overview of the formulation of the many-body problem. The results of the many-body low-energy spectrum are presented in Section \ref{sect:discussion}. Specifically, in Subsection \ref{sect:24holes}, we analyze the case of $N=2$ and $N=4$ holes, while in Subsection \ref{sect:6holes}, we focus on the scenario involving $N=6$ holes. The paper concludes with a summary and key findings in Section \ref{sect:conclusions}.

\section{MODEL}
\label{sect:model}

\subsection{Single Particle Properties}

The single particle  Hamiltonian describing a hole in a single layer of $\text{WSe}_\text{2}$ QD can be expressed as the sum of the bulk Hamiltonian $H_b$ and the confining potential $V_{QD}$  \cite{altintacs2021spin}. In this work, we model the potential $V_{QD}(\vec{r})$ as a Gaussian gate potential given by $V_{QD}(\vec{r})=V_{0}\exp\left(-r^{2}/R_{QD}^2 \right)$, where $V_0$ is the potential depth and $R_{QD}$ corresponds to the QD radius, see Fig. 2(a). Here we will use a typical potential depth $V_0=300$ meV and vary the quantum dot radius $R_{QD}$.  The  wavefunction $|\Phi^{s}\rangle$ of the hole in the QD, associated with the state $s$ can be determined by solving the Schr\"odinger equation:
\begin{equation}
 (H_{b}+V_{QD}(\vec{r})) \ket{\Phi^{s}}= E^s \ket{\Phi^{s}}.
\label{main:hamqdeq1}
\end{equation}
To describe the valence band (VB) wavefunction of our quantum dot, we adopt a method similar to that developed in Ref. \cite{altintacs2021spin} for electrons in the conduction band. In this approach, we construct the VB wavefunction at each wave-vector $k$ as a linear combination of simple Bloch functions on the  tungsten and selenium atoms sublattices. Specifically, we consider six such Bloch functions enumerated by the index $l$ ($l=1,..,6$):
\begin{equation}
\ket{\phi^{\textrm{VB}}_{k\sigma}} = \sum_{l=1}^{6} A^{\textrm{VB}}_{k\sigma, l} \ket{\phi^{\textrm{}}_{k,l}}\otimes\ket{\chi_{\sigma}},
\label{eq:4}
\end{equation}
where $\ket{\chi_{\sigma}}$ represents the spinor part of the wavefunction and
\begin{equation}
\ket{\phi^{\textrm{}}_{k,l}}=\frac{1}{\sqrt{N_{\textrm{UC}}}} \sum_{\vec{R_{l}}=1}^{N_{\textrm{UC}}} e^{i\vec{k}\vec{R_{l}}} \varphi^{}_{l}\left(\vec{r}-\vec{R_{l}}\right),
\end{equation}
are simple Bloch functions built with   orbitals  $\varphi^{}_{l}$ even with respect to the metal plane. $N_{UC}$ is the number of unit cells and $R_l$ defines the position of orbitals in the computational box. By diagonalizing the $6\times 6$ bulk Hamiltonian we obtain the even bulk energy bands $E_{k\sigma}^{\textrm{VB}}$ and wavefunctions $ A^{\textrm{VB}}_{k\sigma,l}$. 

\begin{figure}[ht]
\centering          
\includegraphics[width=9cm, height=7cm]{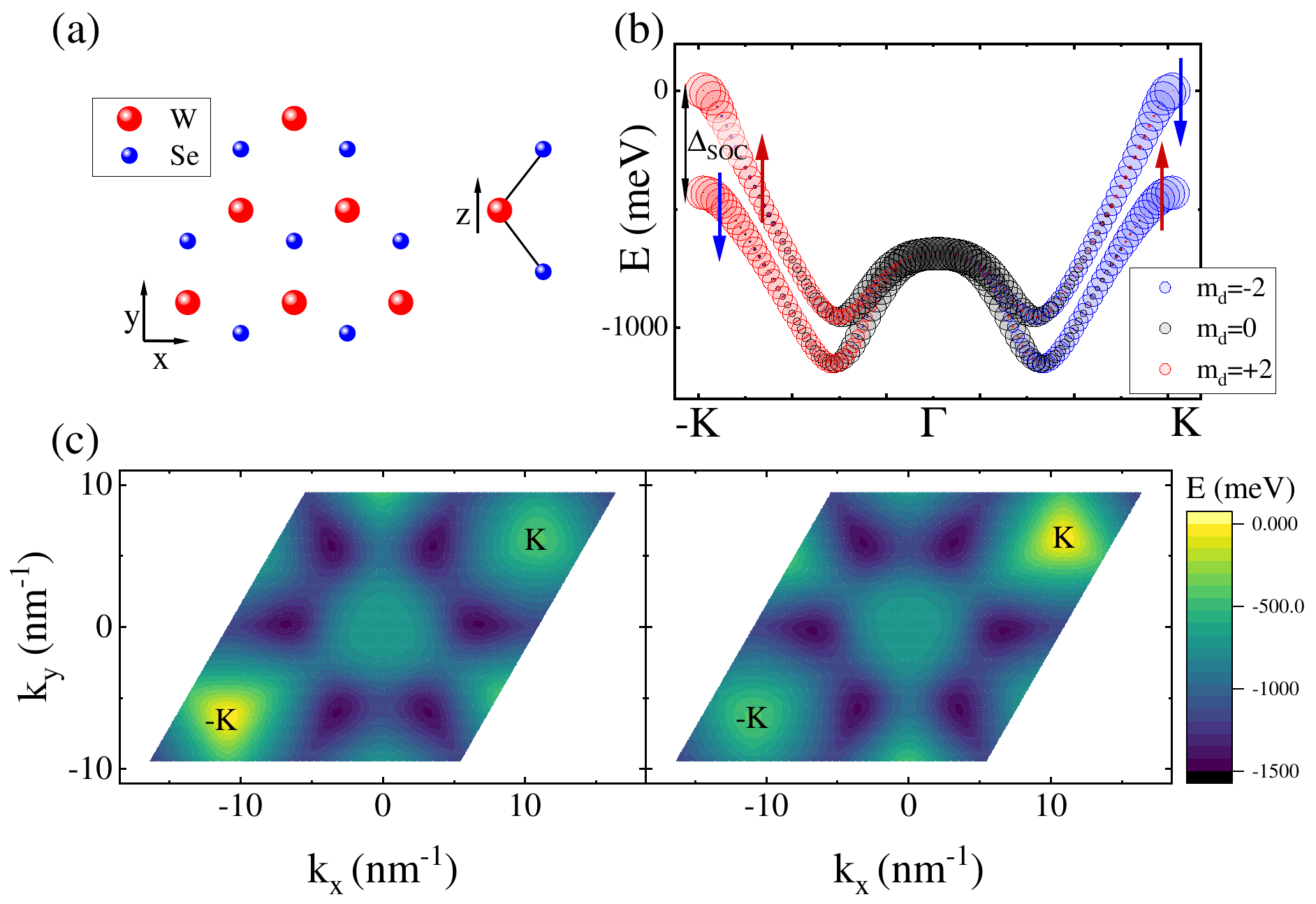}
\caption{
(Color online)  (a) Schematic representation of a monolayer $\text{WSe}_\text{2}$. The red (blue) dots represent W (Se) atoms. (b) Highest energy valence band as a function of wavevector $k$ over the path $-K \rightarrow \Gamma \rightarrow +K $. The  states in the $+K$ valley are composed mainly of $m_{d} = -2$ orbitals, while states in the $-K$ valley are composed mainly of $m_{d} = +2$ orbitals. (c) Energy levels of the highest valence band in the bulk $\text{WSe}_\text{2}$ at different allowed values of $k$ points, shown from left to right for spin-up and spin-down states. The $+K$ and $-K$ points represent the global valley maxima of the valence band. } 
\label{fig1}
\end{figure}
Figure \ref{fig1} shows the energy $E_{k\sigma}^{\textrm{VB}}$ of the hole in the highest even valence band (VB) and provides a map of the valence band energies across the rhombus in the $k$-space where the computations are conducted, encompassing the $+K$ and $-K$ valley maxima. It is important to note that the spin splitting $\Delta_{SOC}$ is opposite in these valleys, with a magnitude of approximately $470$ meV \cite{le2015spin}.

The QD wavefunction $\ket{\Phi^{s}}$ can be expanded in terms of the highest even energy valence band states given by Eq. \ref{eq:4}:
\begin{equation}
\ket{\Phi^{s}}= \sum_{\vec{k}} \sum_{\sigma} B^{\textrm{s,VB}}_{\vec{k}\sigma} \ket{\phi^\textrm{VB}_{\vec{k}\sigma}}.
\label{eq:KetQD}
\end{equation}
The Schr\"odinger equation can now be expressed as an integral equation for the coefficients $B^{\textrm{s,VB}}_{\vec{k}\sigma}$  
\begin{equation}
E_{q\sigma}^{\textrm{VB}} B_{q\sigma}^{\textrm{s,VB}} +\sum_{\vec{k}\sigma'} V_{q,k} A_{q\sigma,k\sigma'}B_{k\sigma'}^{\textrm{s,VB}} = E^s  B_{q\sigma}^{\textrm{s,VB}}.
\label{main:qdeq2}
\end{equation}
Here, the QD confining potential in the wavevector space is represented as a product of  the lateral electrostatic confinement potential $V_{q,k}$ and the band contribution $A_{q\sigma,k\sigma'}$, with
\begin{equation}
\begin{aligned}
V_{q,k}=V_0 \frac{S}{4\pi}R_{QD}^2 \exp\left(-\frac{(k-q)^2 }{4}R_{QD}^2\right),
\end{aligned}
\end{equation}
where $V_{q,k}$ is the Fourier transform of the confining potential. $S$  corresponds to the reciprocal lattice unit cell area. The band structure contribution to the scattering potential $A_{q\sigma,k\sigma'}$ is given by
\begin{equation}
A_{q\sigma,k\sigma'}=\sum_{l} \left(A^{\textrm{VB}}_{q\sigma,l}\right)^{\dagger} \left(A^{\textrm{VB}}_{k\sigma',l}\right).
\end{equation}

To facilitate numerical calculations, we use a computational box that accommodates a maximum of $N_1 \times N_2 = 1011 \times 1011$ unit cells, the specific size depending on the QD defined by metallic gates  under consideration. This corresponds to a total of $3 \cdot N_1 \times N_2 =$ 3,066,363 atoms. To handle such a large number of atoms, periodic boundary conditions are imposed on the computational domain, creating a discrete set of $k$ points in the reciprocal space. Since we focus on the highest valence band states, we apply a cutoff to the $k$ points near $K$ and $-K$ by selecting values of $k$ that satisfy the condition $|k\pm K| < \eta |K|$. This approach helps us to focus on the relevant region of the reciprocal space. To ensure the convergence of the QD single-particle states, we set a fixed value of $\eta = 0.1$. By using this modified set of $k$ points, Eq. \ref{main:qdeq2} becomes computationally manageable, enabling us to perform our calculations effectively.

\begin{figure}[ht]
\includegraphics[width=4.5cm, height=4.2cm]{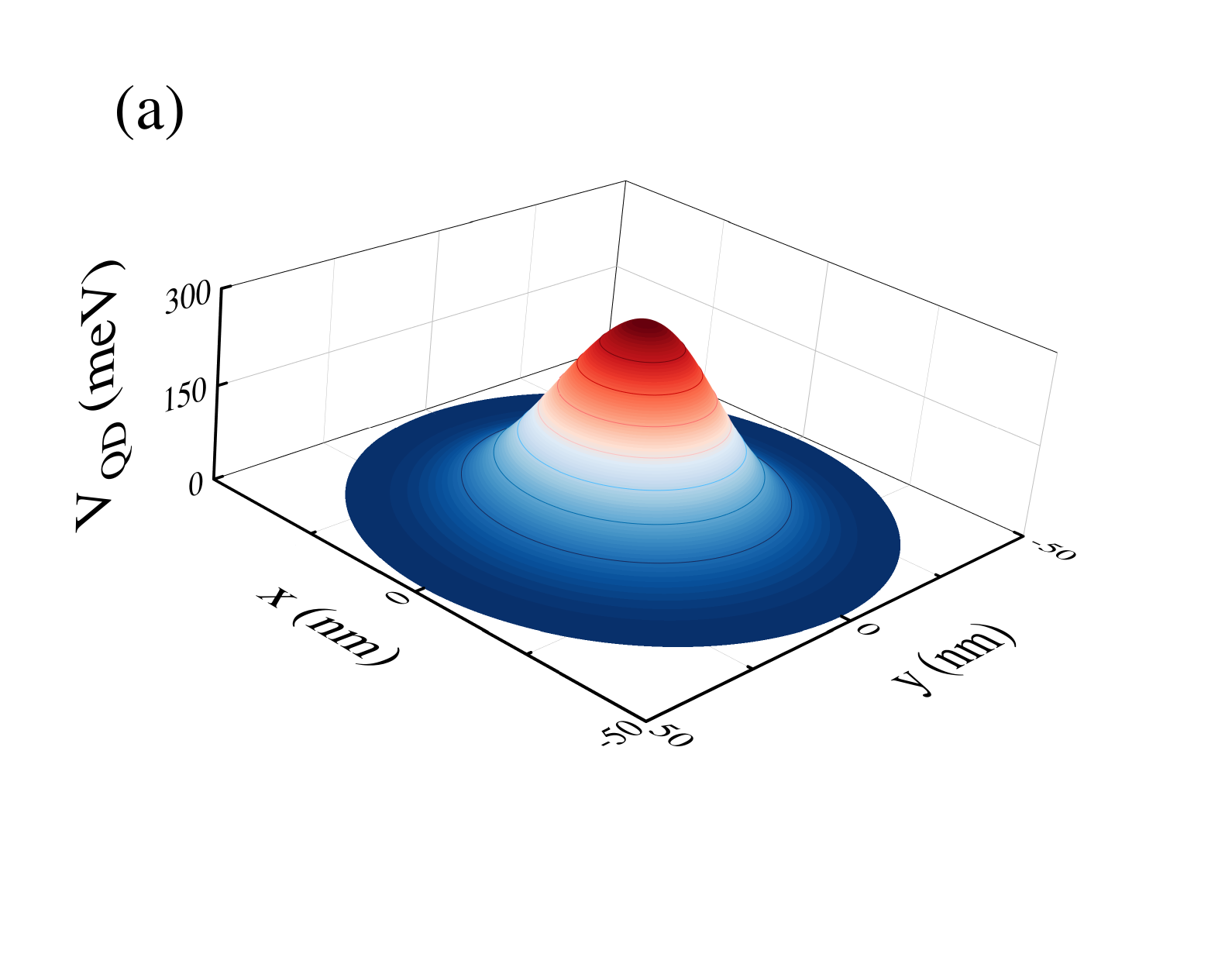}
\hspace{-3em}
\includegraphics[width=4.5cm, height=4.2cm]{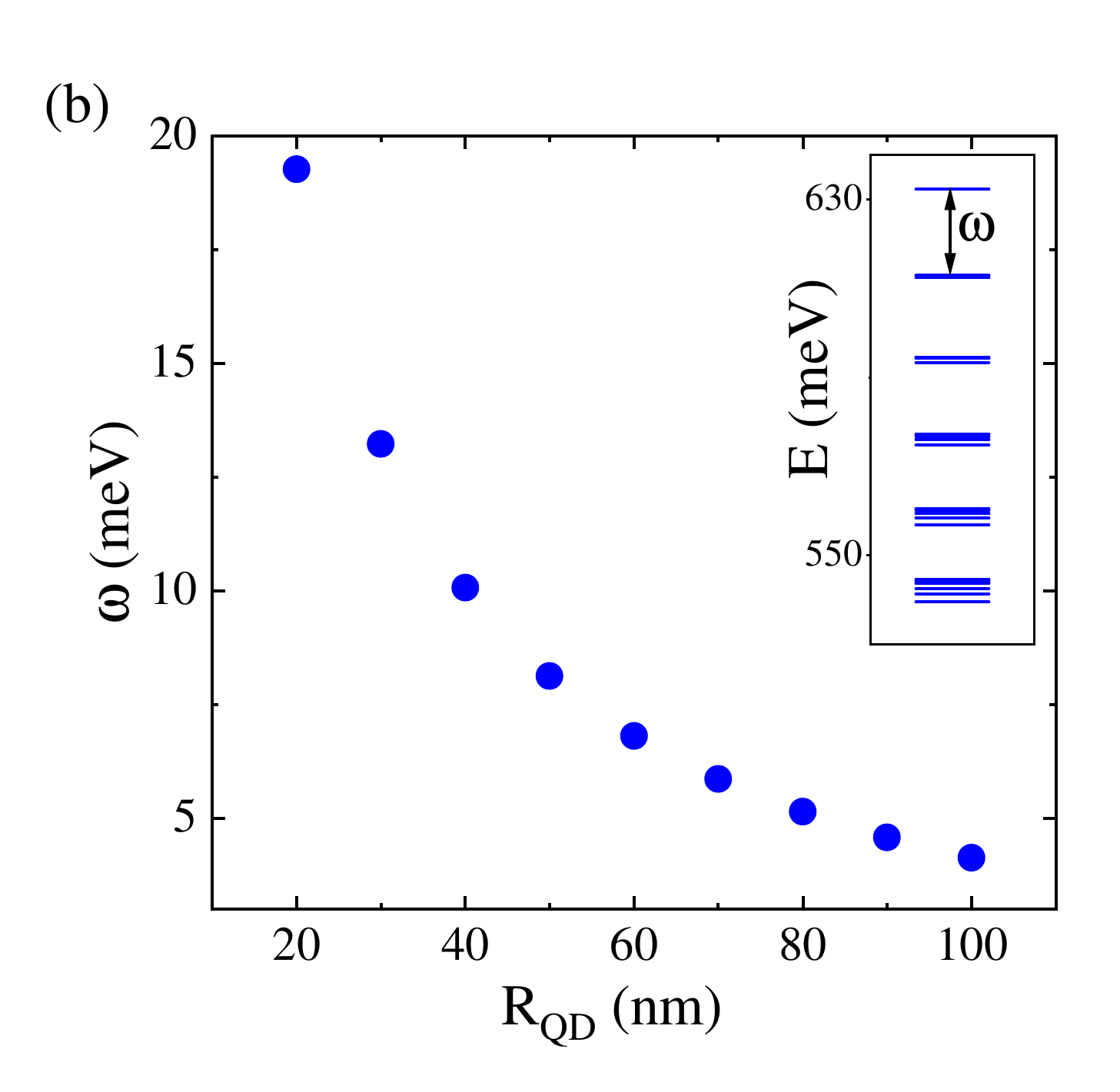}
\caption{
(Color online)  Hole confining potential for a gate potential depth of $V_0 = 300$ meV and a QD radius of $R_{QD}=20$ nm. The potential creates a localized region where holes are confined. (b) Dependence of the  inter-shell spacing $\omega$ on the QD radius. (inset) QD spectrum for $R_{QD}=20$ nm, showing the energy levels of the confined holes. The applied negative gate potential leads to the formation of harmonic oscillator-like shell structures in the states of the QD.}
\label{fig2}
\end{figure}
We obtain the QD energy levels and wavefunctions by solving   Eq. \ref{main:qdeq2}. Figure \ref{fig2} (b) shows the energy difference between the two highest confined shells. The inset in Figure \ref{fig2} (b) shows the  energy levels  of a hole confined within a QD of radius $20$ nm. It is evident from the plot that the energy levels exhibit a distinctive grouping into shells. 
Notably, the highest energy shell comprises of two states,  corresponding to the   spin-down and  spin-up configurations, situated in the  $+K$ and $-K$ valleys, respectively.

Despite the cylindrical symmetry of the confining potential, we observe an intrashell splitting phenomenon. This intriguing observation has been previously reported \cite{bieniek2020effect} and can be compared to self-assembled quantum dots (QDs) subjected to an applied magnetic field \cite{RaymondPRL2004}. The observed intrashell splitting can be attributed to the presence of Berry curvature, which can be understood as an effective magnetic field that acts on the angular momentum states in opposite directions in the K and -K valleys \cite{Zhou2015Berryspliting,Srivastava2015BerrySplit}.

 \subsection{Many-body Interactions}
 
 We now consider the problem where  $N$ holes occupy the valence single-particle states within the QD. In this case, the interacting Hamiltonian for holes can be written as 
\begin{equation}
H=-\sum_{p} E^p c^\dagger_p c_p + \frac{1}{2}\sum_{p,q,r,s} \bra{pq}V\ket{rs} c^\dagger_p c^\dagger_q c_r c_s.
\label{eq:manybodyHam}
\end{equation}
Here, $E^p$  represents the energies of the QD single-particle states, which are obtained by solving  Eq. \ref{main:qdeq2}. The operator $c^\dagger_p $ ($c_p$) creates (annihilates) a hole in the QD state $p$. The Coulomb matrix elements, denoted as $\bra{pq}V\ket{rs}$, are expressed in terms of the QD single-particle orbitals as 

\begin{align}
    \bra{pq}V\ket{rs} =&\int d\vec{r}_{1}\int d\vec{r}_{2}\Phi^{p}(\vec{r}_{1})^{\ast}\Phi^{q}(\vec{r}_{2})^{\ast} V(\vec{r}_{2}-\vec{r}_{1})\nonumber\\
    &\times \Phi^{r}(\vec{r}_{2})\Phi^{s}(\vec{r}_{1}),
\end{align}
where $\Phi^{p}(\vec{r})=\bra{\vec{r}}\ket{\Phi^{p}}$. The  interaction $V$ is given by the Coulomb potential with Keldysh screening \cite{Keldysh1979,LudkaQD2020} 
\begin{align}
    V(\vec{r}_{2}-\vec{r}_{1})= \nonumber &\frac{e^2}{4\pi\epsilon_0\epsilon}\frac{1}{(2\pi)^2}\int  \frac{2\pi}{|k|}\frac{1}{1+2\pi\alpha |k|}e^{-|z_2-z_1||\vec{k}|}\\ 
    & \times e^{i\vec{k}\cdot(\vec{\rho}_2-\vec{\rho}_1)} d^2 \vec{k},
\end{align}
where $\vec{r}=(\vec{\rho},z)$, $\alpha=2.2$ is the 2D polarizability and $\epsilon$ is the relative dielectric constant. We will use  $\epsilon =6.0$ in this work. 

To diagonalize the many-body  Hamiltonian represented by Eq. \ref{eq:manybodyHam} we use the configuration-interaction (CI) approach. This involves generating all  possible configurations of the $N$ holes occupying a specific  number of  single-particle orbitals, constructing the many-body Hamiltonian matrix in the space of configurations, diagonalizing this matrix, and obtaining the energy spectrum and many-body eigenstates. 

\section{Numerical exact diagonalization and Discussion of results}

\label{sect:discussion}

We consider QDs of diameters up to $100$ nm and up to $42$ single-particle valence band states (6 shells), which we then fill with up to 6 holes. Due to the large spin-orbit coupling in the valence band of  $\text{WSe}_\text{2}$, the complete shell of single-particle levels splits into sub-shells. This spin splitting gives rise to a spin-valley locking effect, whereby the QD states near the top of the valence band have spin down and up in the $+K$ and $-K$ valleys, respectively. Hence, it is safe to concentrate only on the highest sub-shells that are well isolated from the sub-shells pushed deep into the valence band, as shown in Figure \ref{fig1}. 

\begin{figure}[ht]
\includegraphics[width=9cm, height=5cm]{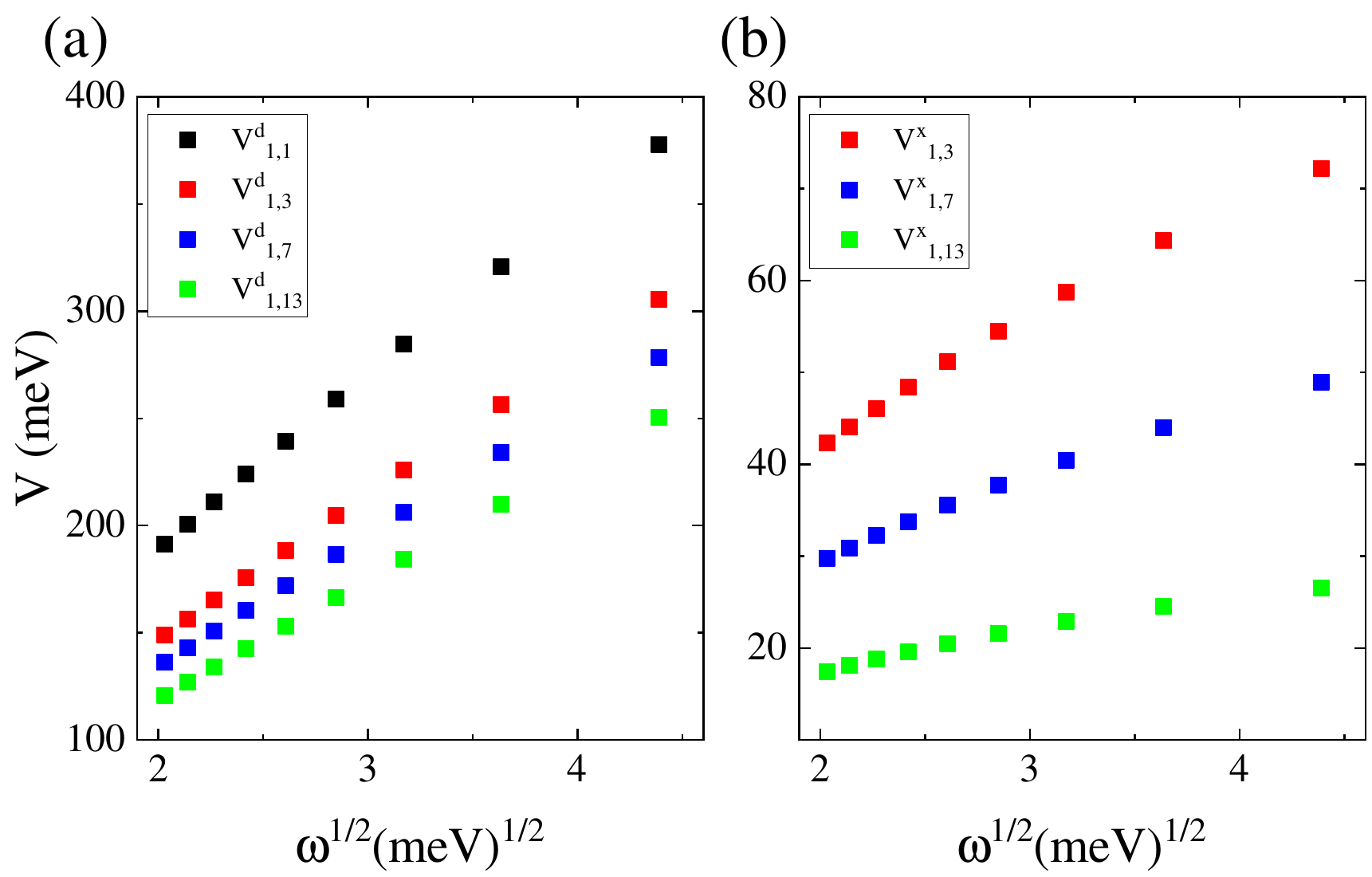}
\caption{(Color online) Coulomb matrix elements for single particle QD states as a  function of shell energy spacing $\sqrt{\omega}$. (a) Direct Coulomb matrix elements and (b) exchange Coulomb matrix elements with Keldysh screening. The notation $V^d_{i,j}$ ($V^x_{i,j}$) denotes the interaction between the  levels $i$ and $j$, indicating the strength of the direct (exchange) interaction between these levels. }
\label{fig3}
\end{figure}

Shell spacing of the energy levels is inversely proportional to the QD's radius, which can be seen in Figure \ref{fig2} (b). On the other hand, the strength of the  Coulomb interactions is directly proportional to the square root of the shell spacing $\omega$ which can be seen in Figure \ref{fig3}. Here, $V^d_{i,j}$ ($V^x_{i,j}$)  refers to  the direct (exchange) interaction between the  levels $i$ and $j$. Furthermore, the effect of the interactions is increasing with the QD radius as shown in Figure  \ref{fig3}.

\subsection{$N=2,4$ holes }
\label{sect:24holes}
\begin{figure}[ht]
\includegraphics[width=9cm, height=4.5cm]{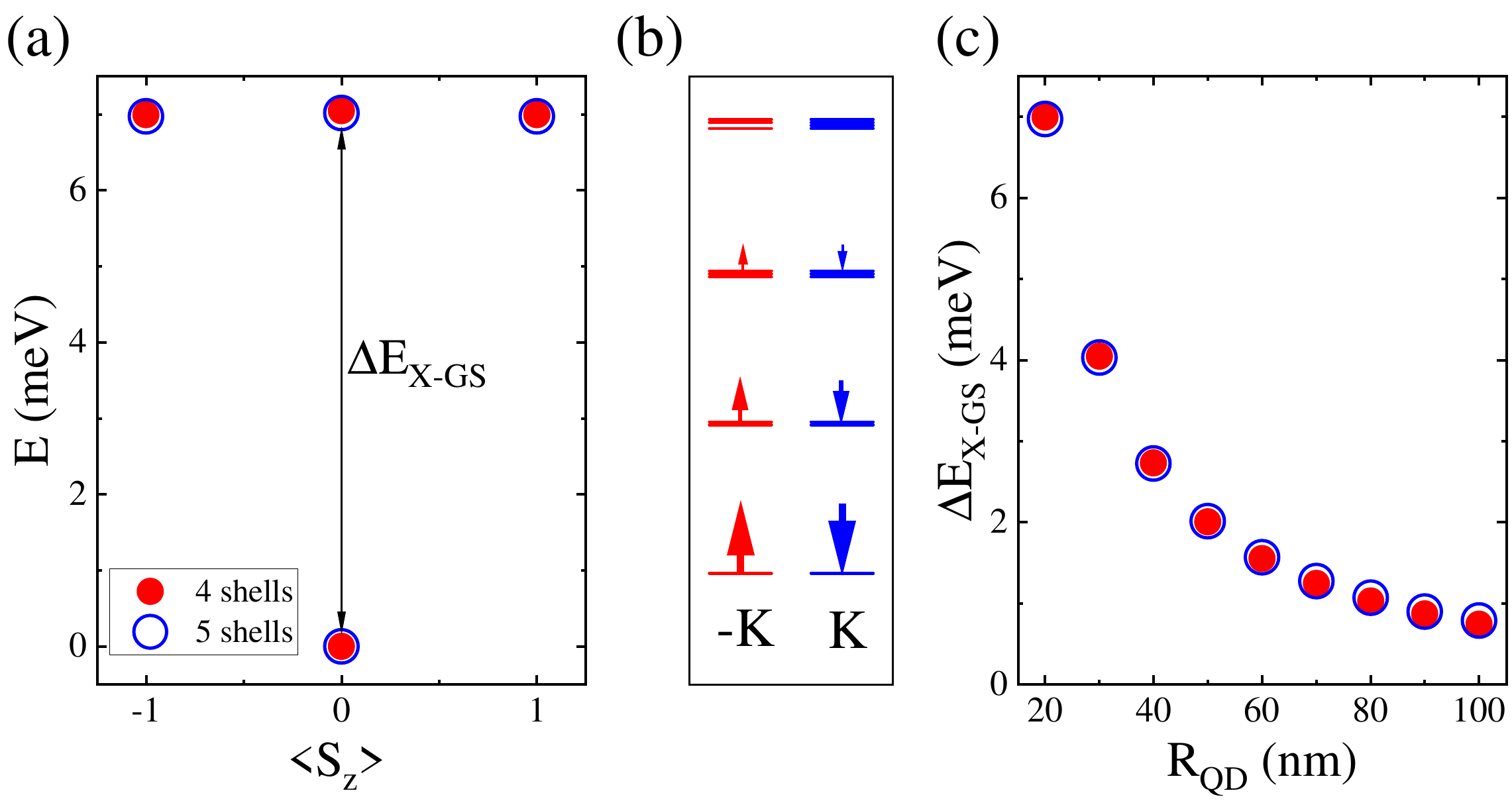}
\caption{
(Color online) (a) Low energy spectrum for $N = 2$  holes in QD with $R_{QD} = 20$ nm QD. The ground state is a singlet, followed by triplet excited states. (b) Schematic representation of the ground-state occupation of single-particle QD orbitals. (c) Dependence of the singlet-triplet gap on the QD size, showing how energy separation between singlet and triplet states varies with the QD radius. 
}
\label{2holes}
\end{figure}

Figure \ref{2holes} (a) shows the low energy spectrum for a QD with $R_{QD}=20$ nm. Ground state reference energy is chosen at $E=0$. The ground state is an unpolarized and fully intervalley antiferromagnetic state. 

The stability of the ground state compared with other phases is partly determined by the energy difference between the ground state and the first excited state $\Delta E_{X-GS}$. Figure \ref{2holes} (c) shows the dependence of the singlet-triplet gap on the QD size. In this case, by increasing the radius (increasing interaction effects) the gap decreases. 

For $N=2$, the non-interacting ground state has the first shell completely filled. However, the Coulomb  interaction changes this picture notably. Figure \ref{2holes} (b) schematically shows the occupation of the single particle QD states in the many body ground state. The holes tend to occupy mainly the first shell like in the non-interacting ground state, although  there is some finite occupation in the third shell induced by the Coulomb repulsion. In this case, the result for 4 and 5 shell shells are equivalents for all the QD sizes considered.

\begin{figure}[ht]
\includegraphics[scale=0.16]{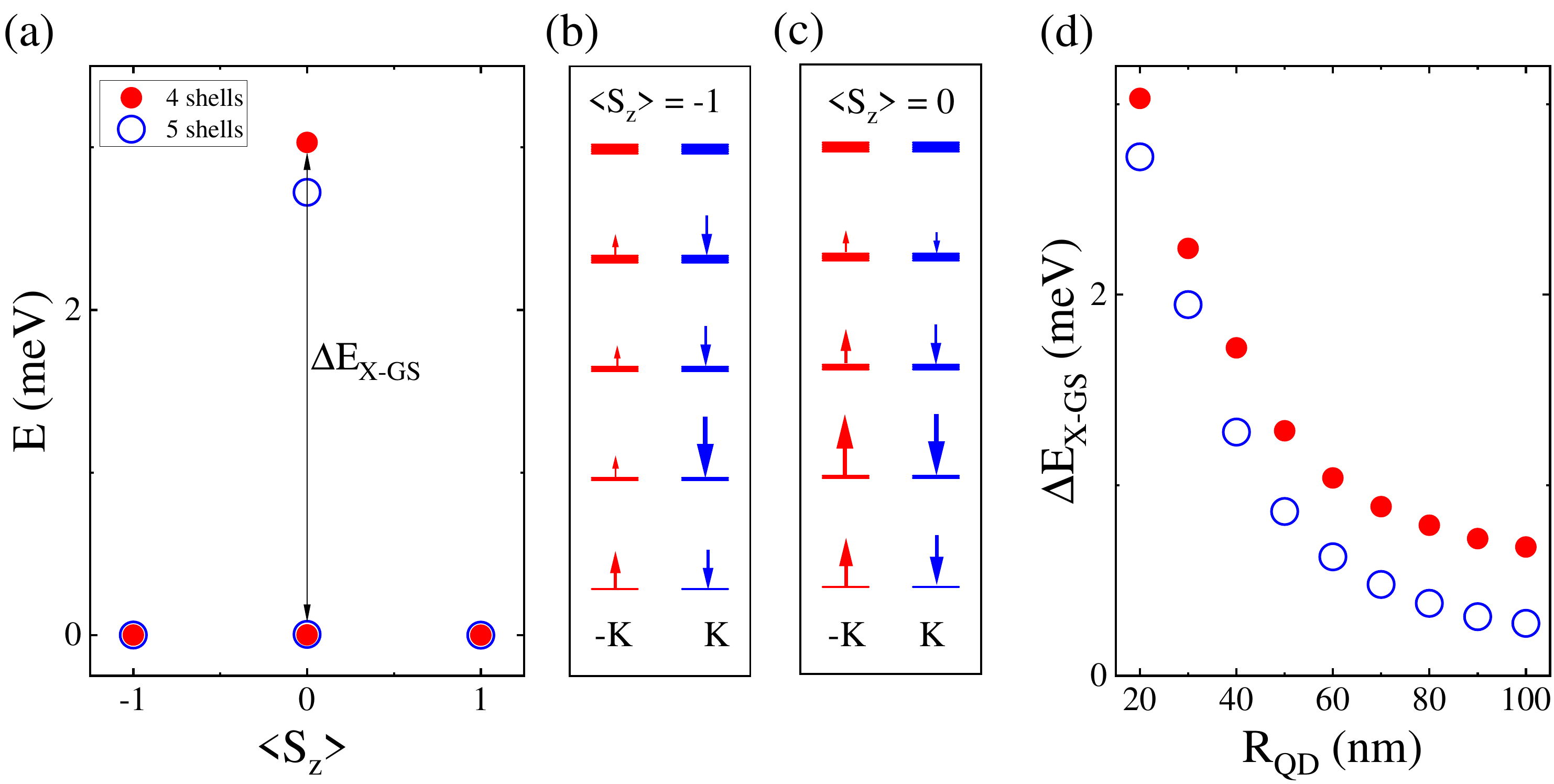}
\caption{
(Color online) (a) Low energy spectrum for $N = 4$  holes in QD with $R_{QD} = 20$ nm QD. The ground state is a triplet, followed by a singlet excited state. (b), (c) shows the schematic representation of the ground-state occupation of single-particle QD orbitals. (d) Dependence of the triplet-singlet gap with the QD size.}
\label{4holes}
\end{figure}

For $N=4$ the non-interacting ground state has the second shell half-filled. The low energy spectrum for $N=4$ interacting holes is shown in Figure \ref{4holes}. Here, the excitation gap $\Delta E_{X-GS}$ is smaller than for $N=2$ holes, the ground state is a triplet,  with two states spin and valley polarized, and one unpolarized. Figure \ref{4holes} (d) shows the dependence of the triplet-singlet gap on the QD size. The difference in the  gap between the result for 4 and 5 shells is almost constant over the range of QD radius considered. 

Figures \ref{4holes} (b) and (c) provide a schematic illustration of the ground-state occupation of single-particle QD orbitals for both unpolarized ($S_z = 0$) and polarized ($S_z = 1$) states, respectively. In the unpolarized case, the occupation pattern is similar to that of non-interacting systems, except for some higher energy orbitals being occupied due to the interactions. However, in the polarized state, the occupation profile deviates more  significantly from the non-interacting case. Specifically, the first shell is partially spin-polarized, while the second shell is fully spin-polarized. This finding highlights the strong impact of the electron-electron interactions on the ground-state properties of QDs, especially in the polarized regime.

\subsection{$N=6$ holes states}
\label{sect:6holes}

\begin{figure*}[ht]
\includegraphics[scale=0.18]{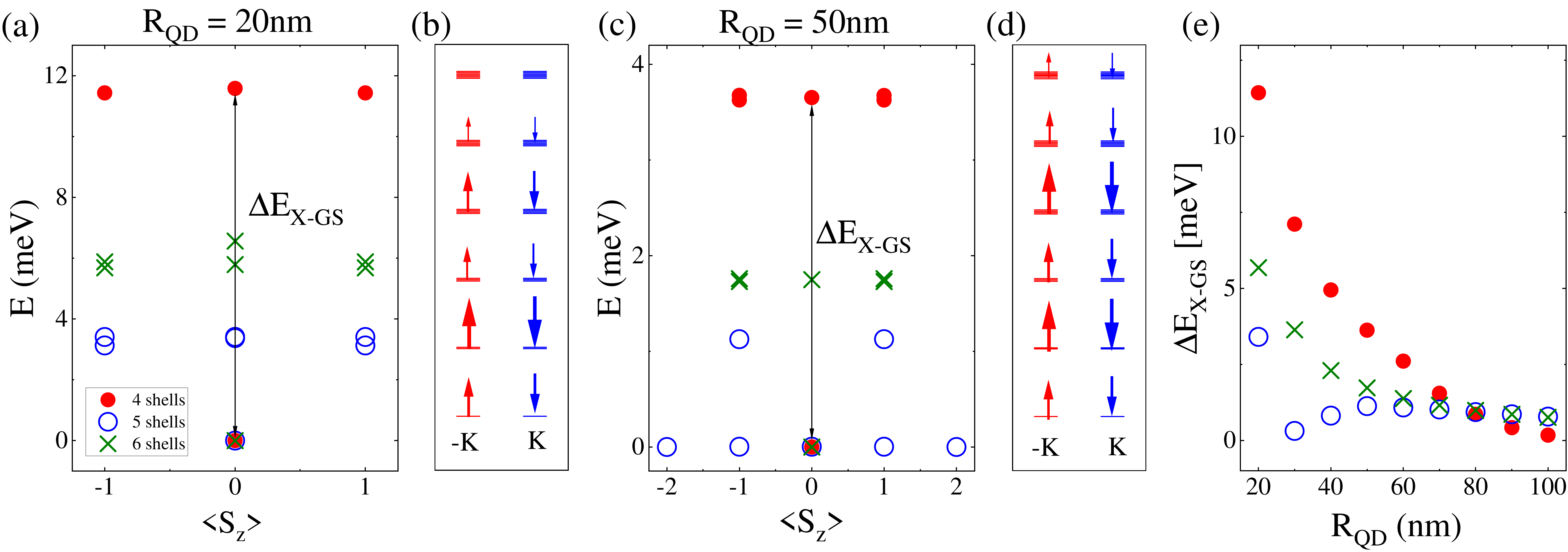}
\caption{
(Color online) (a) Low-energy spectrum for $N=6$ holes in a QD with $R_{QD} = 20$ nm. (b), (d) Schematic representation of the ground-state occupation of single-particle QD orbitals. (c) Low-energy spectrum for $N=6$ holes in a QD with $R_{QD} = 50$ nm. (e) Dependence of the excitation gap on QD size for different numbers of shells included in the many-body calculation. The excitation gap represents the energy separation between the ground state  and the first excited states. }
\label{6holes}
\end{figure*}

Figure \ref{6holes} (a) depicts the low-energy spectrum for a QD with $R_{QD}=20$ nm, considering the presence of $N=6$ holes. The corresponding spectrum for $R_{QD}=50$ nm is shown in Fig. 6 (c). The behavior observed in the spectrum is complex and depends on the number of shells considered in the CI calculations. The number of shells is controlled by potential depth $V_0$.

For calculations involving $4$ shells, the ground state is consistently a singlet across different values of $R_{QD}$. However, the excitation gap  decreases as the QD radius  increases.  In the case of $5$ shells included in the basis, the ground state remains  a singlet for $R_{QD} < 30$ nm, while for  $R_{QD} > 30$ nm the interaction effects become more significant, leading to a ground state with a quintuple character. To  further explore this phenomenon, an additional shell is included  in the calculations for $N=6$ holes. Surprisingly, the resulting low-energy spectrum exhibits more similarities to the result obtained using only $4$ shells rather than those obtained with $5$ shells. Similar behavior has been observed in previous studies of parabolic QDs  \cite{MarekShells2003}.

This behavior can be attributed to the repulsion between holes, particularly the direct term of the Coulomb interaction. As depicted in  Figures \ref{6holes} (b) and (d), by increasing $R_{QD}$ the particles tend to spread into more shells, thereby minimizing the Coulomb repulsion. However, this also results in an increase in the kinetic energy. The ratio between the Coulomb interaction and the kinetic inter-shell gap ($\omega$) increases with the QD radius (see Figure  \ref{fig3}). This implies that the kinetic energy gain associated with occupying higher energy shells is only compensated for larger QD sizes.

To obtain a more accurate determination of the ground state for $N=6$ holes, irrespective of the depth of confining potential, it is necessary to include additional shells in our calculations. However, we have already reached the limits of our CI capabilities in terms of computational resources. One potential approach to address this challenge is to employ a ground state calculation using a Matrix Product State (MPS) - Density Matrix Renormalization Group (DMRG) scheme \cite{WhitePRL92,SchollwockReview2005}. This method while variational in nature has proven successful in studying many-body systems and can potentially extend our results to a larger number of shells. However, the presence of long-range interactions, as observed in Figure \ref{fig3}, may complicate the application of the MPS-DMRG scheme for this particular system. In future work, we plan to investigate different  many-body states by extending the number of shells and exploring the feasibility and benefits of the MPS-DMRG approaches to obtain more comprehensive insights into the system's behavior. However, we also point to the fact that the number of shells is controlled by the depth $V_0$ of the confining potential and hence different ground states for different depths predicted  in this study are to be expected. 

\section{Conclusions}
\label{sect:conclusions}
In this study, we presented  an investigation of the electronic properties of finite numbers of valence holes in gated $\text{WSe}_\text{2}$ quantum dots, considering the influence of spin, valley, orbital, and many-body interactions on the low-energy spectrum. By employing a multi-million atom tight-binding model solved in the wave-vector space, we have obtained single-particle states characterized by the harmonic oscillator-like states with a shell spacing $\omega$, which depends on the size of the QD and the depth of the gate potential. We have extended our study to include many-body interactions for systems consisting of up to $N=6$ holes. The significant spin-orbit splitting in the valence band has led to ground state configurations with spin-down in the $+K$ valley and spin-up in the $-K$ valley, demonstrating the spin-valley locking effect. Our results emphasize the importance of the many-body interactions in controlling the behavior of these systems. Furthermore, for $N=6$ holes and small QD radii ($R_{QD}<30$ nm), we have found that the ground states can be accurately determined using around 30 single-particle states. However, for larger QD radii ($R_{QD}>30$ nm), where the Coulomb interaction becomes stronger, it is necessary to include a larger number of single-particle states to properly establish the ground state. This highlights the sensitivity of the system's behavior to the interplay between confinement and many-body effects.  

\section*{Acknowledgments}
The authors thank H. Allami, M. Cygorek,  J. Manalo,  L. Szulakowska, Y. Saleem, M. Mohseni,  J. Boddison-Chouinard, A. Bogan, A. Luican-Mayer and L. Gaudreau for valuable discussions. 

This work
was supported by the  Quantum Sensors Challenge Program at the National Research Council of
Canada,  NSERC QC2DM Strategic Grant No. STPG-
521420, NSERC Discovery Grant No. RGPIN- 2019-05714, and University of Ottawa Research Chair in Quantum Theory of Materials, Nanostructures, and Devices. M.B. acknowledges financial support from the Polish National Agency for Academic Exchange (NAWA), Poland, grant PPI/APM/2019/1/00085/U/00001. This research was enabled in part by support provided by the Digital Research Alliance of Canada (alliancecan.ca).


\bibliography{refs}

\end{document}